\documentclass[10pt]{article}

\usepackage{cite}
\usepackage{amsmath,amssymb,amsfonts}
\usepackage{algorithmic}
\usepackage{algorithm}
\usepackage{graphicx}
\usepackage{textcomp}
\usepackage{xcolor}
\def\BibTeX{{\rm B\kern-.05em{\sc i\kern-.025em b}\kern-.08em
    T\kern-.1667em\lower.7ex\hbox{E}\kern-.125emX}}

\usepackage{listings}
\usepackage{multirow}
\usepackage{array}
\usepackage[margin=1in]{geometry}
\usepackage{rotating}
\usepackage{hyperref}

\newtheorem{definition}{Definition}

\newcommand{\AC}{\lambda}
\newcommand{\AR}{\alpha}
\newcommand{\BC}{\beta}
\newcommand{\block}[2]{#1_{#2}}
\newcommand{\blockSize}{\emph{block size}}

\begin{document}

\title{Open-source Stand-Alone Versatile Tensor Accelerator.}

\author{Anthony Faure-Gignoux$^1$ \and Kevin Delmas$^2$ \and Adrien Gauffriau$^1$ \and Claire Pagetti$^2$ }
\date{%
    $^1$\textit{Airbus}, France\\%
    $^2$\textit{ONERA}, France\\[2ex]%
    \today
}

\maketitle

\begin{abstract}
Machine Learning (ML) applications demand significant computational resources, posing challenges for safety-critical domains like aeronautics. 
The Versatile Tensor Accelerator (VTA) is a promising FPGA-based solution, but its adoption was hindered by its dependency on the TVM compiler and by other code non-compliant with certification requirements.
This paper presents an open-source, stand-alone Python compiler pipeline for the VTA, developed from scratch and designed with certification requirements, modularity, and extensibility in mind.
The compiler’s effectiveness is demonstrated by compiling and executing LeNet-5 Convolutional Neural Network (CNN) using the VTA simulators, and preliminary results indicate a strong potential for scaling its capabilities to larger CNN architectures.
All contributions are publicly available.
\end{abstract}

\section{Introduction}
Machine Learning (ML) applications are increasingly demanding significant computational resources.
This is particularly challenging in domains constrained by hardware limitations, such as aeronautics. 
Multi-core CPUs often struggle to meet these demands efficiently.
Therefore, dedicated hardware accelerators, offering highly parallel computing resources, are a way forward.

FPGAs are widely used in aeronautics due to their reconfigurability and potential for tailored, high-performance designs.
Therefore, FPGA-based accelerators are a promising solution for executing ML in aeronautics.
In this context, the Versatile Tensor Accelerator (VTA)~\cite{VTA_paper} emerges as a compelling open-source solution.
Designed specifically for the efficient execution of matrix multiplication operations, its architecture is inherently flexible and configurable, allowing it to be tailored to various target FPGA devices.
Furthermore, the use of a VTA-inspired accelerator by the aeronautical company Daedalean for running its Convolutional Neural Networks (CNNs)~\cite{daedalean} highlights the platform's potential as an industrial-grade solution.
Finally, the VTA incorporates both a High-Level Synthesis (HLS)~\cite{hls} description, for FPGA implementation on an UltraScale+, and a CHISEL~\cite{chisel} description, for cycle-accurate simulation.
A second simulator in C++ provides a functional simulation of the VTA behaviour.

However, the practical adoption of VTA faces several substantial challenges, 
as it must adhere to the established avionics guidelines for hardware (DO-254~\cite{DO254}) and software (DO-178C~\cite{DO178C}).
The VTA was highly dependent on the TVM compiler~\cite{tvm}, which is not compliant with aeronautics regulations~\cite{acetone}. 
Indeed, the TVM compiler performs numerous transformations, making it complex to ensure traceability; it also prevents fine-grained user control over the compilation chain. 
Lastly, the TVM project chose to deprecate the VTA.

Consequently, we provide an alternative open-source compilation chain for VTA and a live GitHub repository of associated tools (e.g., simulators), all developed in consideration of the applicable certification documents.
An in-depth investigation of the VTA was necessary to accurately understand its architecture, especially the memory allocation done by the TVM compiler (Section~\ref{sec:background}). 
This investigation enabled us to develop from scratch a compilation chain for the VTA.
First, we delve into matrix multiplication compilation, which is key to executing CNNs on the VTA (Section~\ref{sec:compiler}).
Second, we highlight the link between matrix multiplications and CNN execution; then, we extend the compiler to CNN compilation (Section~\ref{sec:CNN}).
This compiler aims to meet certification requirements. 
It provides a simpler and more transparent toolchain, enabling users to trace the various transformations from the CNNs to the VTA matrices.
It also offers a static implementation strategy enabling a more predictable execution time.
Additionally, the modular design of the compiler allows using and testing different implementation strategies. 
Finally, we compiled the LeNet-5 CNN with our compiler and applied it to the VTA simulators to highlight its efficiency (Section~\ref{sec:simulation}). 
All our work is open-source and available via GitHub~\cite{onera_github}.
This work enhances VTA’s applicability for safety-critical systems by offering a verifiable and independent compilation pathway.

\section{Background: the Versatile Tensor Accelerator}
\label{sec:background}
This section reviews the existing VTA, detailing its established architecture, functional behaviour, and memory management principles, as originally implemented within the TVM compiler framework.

\subsection{The VTA architecture}
\label{sec:VTA_architecture}
\begin{figure}[htbp]
    \centering
    \includegraphics[width=.7\linewidth]{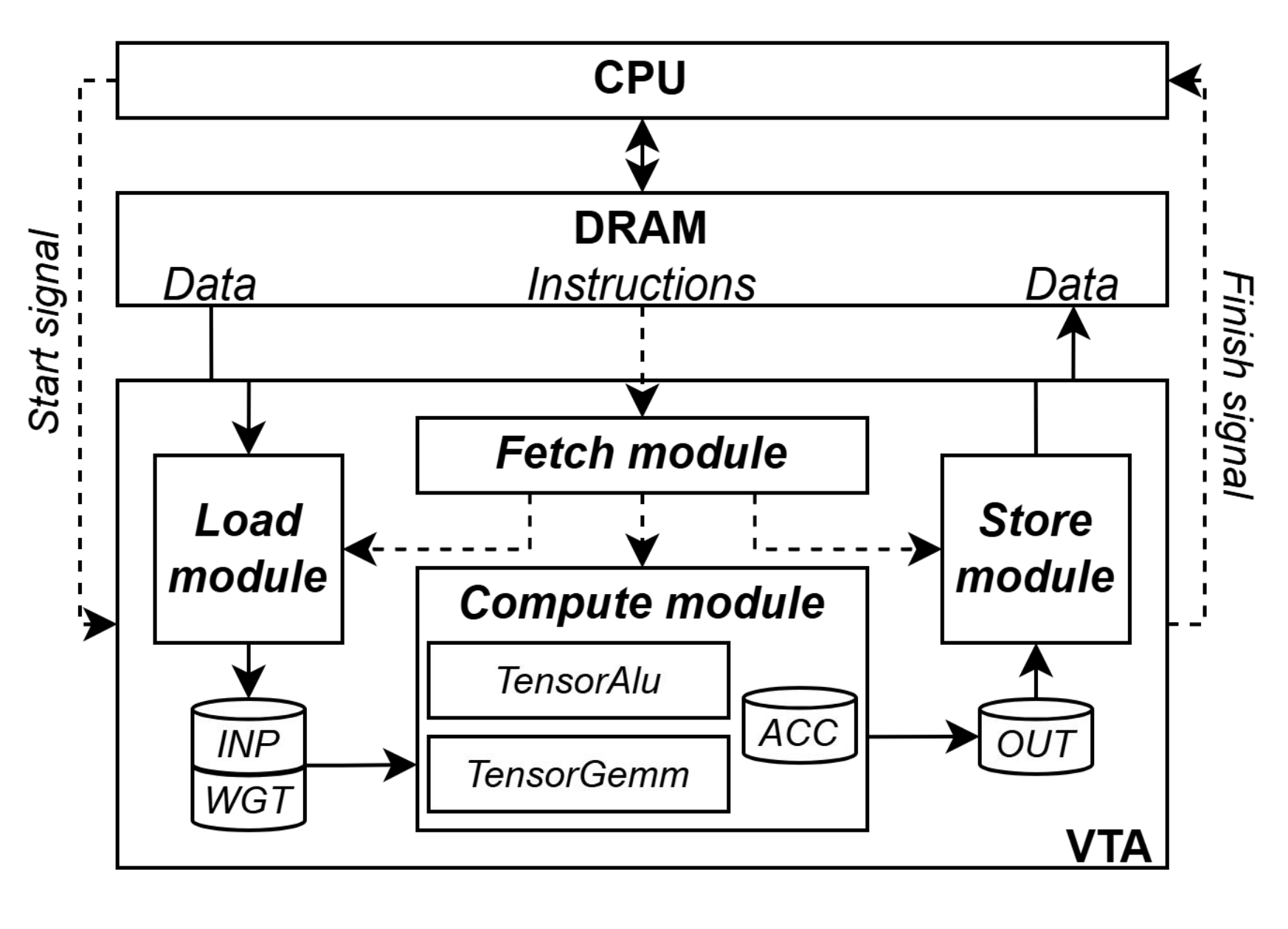}
    \caption{Hybrid architecture}
    \label{fig:highlevel_vta}
\end{figure}

\textbf{Hybrid architecture:}
A hybrid architecture consists of general-purpose CPUs and accelerators, which are highly specialised hardware designed to process large volumes of data efficiently.
The VTA is an accelerator intended for use within such hybrid architectures under the control of a CPU, as shown in Figure~\ref{fig:highlevel_vta}. 
Practically, the CPU sends a start signal containing the memory address of the instructions.
The VTA loads the instructions from the DRAM. 
Within the instructions, it reads the DRAM addresses of the data.
The VTA locally copies the data and then performs its operations on it. 
The result is written back to the DRAM.
Finally, the VTA sends a completion signal to the CPU.
It can then manipulate the data from the DRAM and initiate a new VTA execution.

\textbf{VTA architecture:}
The VTA contains a local SRAM memory segmented into four buffers: input (INP), weight (WGT), accumulator (ACC), and output (OUT). 
INP, WGT, and OUT data are encoded as signed 8-bit integers, whereas ACC uses signed 32-bit integers. 
The VTA manipulates data structures rather than individual data elements. 
INP, ACC, and OUT are vectors with \blockSize~elements, whereas WGT is a square matrix of \blockSize~$\times$ \blockSize~elements. 
The \blockSize~value is a VTA hardware configuration parameter, and its default value is 16, i.e., an INP vector contains 16 elements.
The SRAM uses logical addresses to reference these data structures.

The VTA is also composed of four modules.
The Fetch module dispatches the instructions to the three other modules that operate in parallel.
The Compute module is further subdivided into two submodules: TensorGemm, responsible for vector-matrix multiplications, and TensorAlu, which handles element-wise operations (e.g., ReLU – Rectified Linear Unit). 

TensorGemm performs multiplications between INP vectors and WGT matrices.
The resulting products are then accumulated in ACC vectors.
TensorAlu applies element-wise operations to the ACC vectors.
Finally, the OUT vectors are written to the DRAM.
It is important to note that OUT vectors are truncated ACC vectors.

\subsection{DRAM data mapping} \label{sec:data_mapping}
\textbf{Logical addresses:}
DRAM is a large global storage accessible by both the CPU and the VTA.
Usually, the memory is segmented into regions, with each region allocated to a specific application. 
In our case, a region of memory within the DRAM is assigned to the VTA to store its data and instructions.
The memory address of this region is $\mathit{offset}$.

The DRAM memory system uses both physical and logical addresses. 
The logical addresses serve as an abstraction of the physical hardware addresses.
The VTA manipulates logical addresses computed from the physical addresses as defined in Definition~\ref{def:phy_log_addr}.

\begin{definition}[Physical and logical address relation]
    \label{def:phy_log_addr}
    \itshape
    Let $\mathit{phy\_addr}$ be the physical DRAM address, $\mathit{log\_addr}$ be the logical VTA address, $\mathit{offset}$ be the memory offset of the DRAM region allocated to the VTA, $\mathit{precision}$ be the data representation in Bytes (e.g., 1 Byte for an INP and 4 Bytes for an ACC), $\mathit{nb\_elem}$ be the number of elements in the structure (e.g., \blockSize~for the vectors and \blockSize~$\times$ \blockSize~for the matrices).
    The relation between $\mathit{phy\_addr}$ and $\mathit{log\_addr}$ is as follows:
    \[ \mathit{log\_addr} =  \Big\lfloor \frac{\mathit{phy\_addr} - \mathit{offset}}{\mathit{precision} \times \mathit{nb\_elem}}  \Big\rfloor  \]
\end{definition}

This mechanism facilitates direct referencing of a vector or a matrix as opposed to its individual elements.
For example, the two consecutive logical addresses $\mathit{inp\_addr}$ and $\mathit{inp\_addr}+1$ correspond to the INP vectors $n$ and $n+1$. 
In this paper, the addresses, whether physical or logical, are denoted by ”@” followed by a hexadecimal value.

\textbf{DRAM allocation:}
The TVM compiler is based on a memory allocation function that oversees the management of DRAM memory in 4~KiB pages.
The TVM compiler allocates contiguous physical addresses and provides the corresponding logical addresses.
Each memory allocation begins on a new page, even if the previous page is only partially filled.
The TVM compiler opts for distinct allocation of the various data types, allocating first the INP vectors, then the WGT matrices, and so on.

\begin{figure}[htbp]
    \centering
    \includegraphics[width=.7\linewidth]{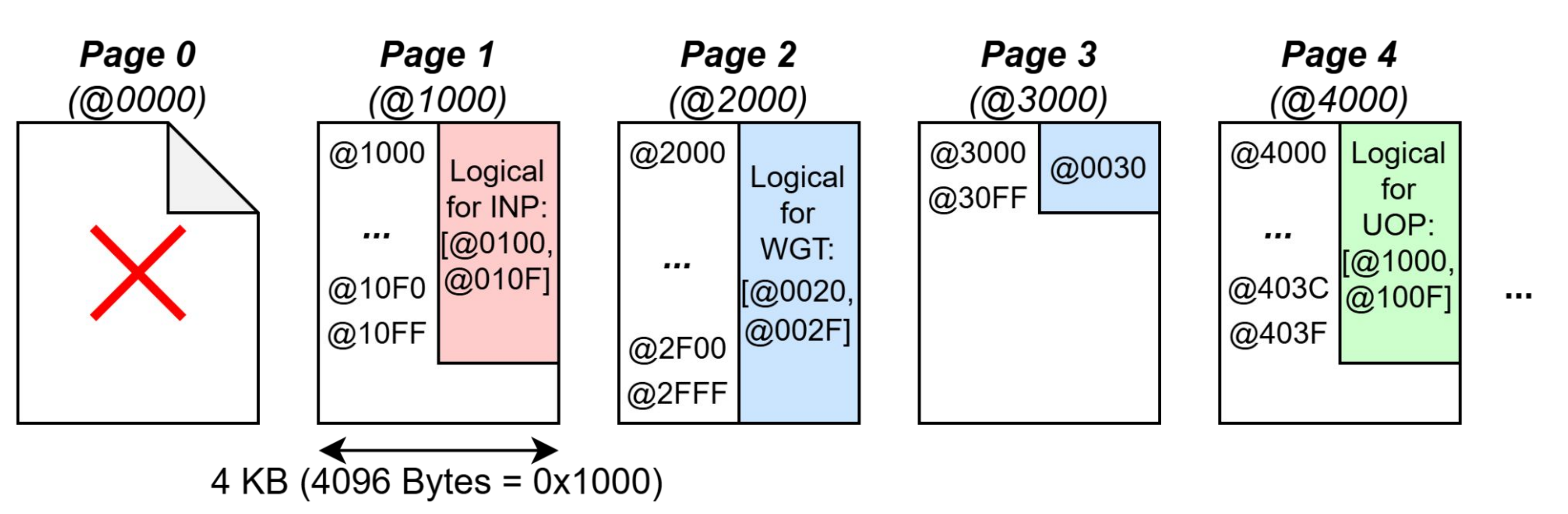}
    \caption{DRAM allocation example}
    \label{fig:dram}
\end{figure}

Figure~\ref{fig:dram} illustrates an example of memory allocation with the memory offset set to $\mathit{offset}=0$.
Initially, the pointer is on Page~0 ($\mathit{phy\_addr} =$ @0000).
The TVM compiler allocates 256~Bytes.
It first advances the pointer to the start of the next available page, Page~1 ($\mathit{phy\_addr} =$ @1000). 
Then it allocates the contiguous physical addresses @1000 to @10FF.
The pointer is now positioned at $\mathit{phy\_addr} =$ @1100.
Next, the TVM compiler allocates 4352~Bytes, exceeding the capacity of a single 4~KiB page. 
The pointer is advanced to Page~2 ($\mathit{phy\_addr} =$ @2000), and the requested bytes are allocated using the physical addresses @2000 to @30FF. 
The pointer is consequently positioned at @3100.

The DRAM allocation mechanism reserves contiguous regions for specific data structures.
For instance, the second allocated region shown in Figure~\ref{fig:dram} is designated for 17 WGT matrices.
Each matrix has a size of 256~Bytes.
The logical address of each WGT matrix can be deduced from Definition~\ref{def:phy_log_addr}. 
The logical address of the first WGT matrix is: 
$\mathit{log\_addr} =$ @2000 $/ 256 =$ @0020.

\subsection{Instructions}
\label{sec:VTA_instructions}
The VTA uses a Complex Instruction Set Computer (CISC) architecture. The host CPU generates 128-bit instructions.
Instructions operate using logical addresses.
There are three distinct types of instructions: (i) LOAD / STORE, (ii) GEMM, and (iii) ALU.
All instructions share two common fields: a 3-bit OPCODE and a 4-bit DEPT\_FLAG.
The OPCODE specifies the type of instruction, while the DEPT\_FLAG, corresponding to the dependency flags, manages resources shared among the modules.
The VTA is designed primarily for accelerating matrix multiplications; consequently, its GeMM instructions, executed by the TensorGemm module to perform these fundamental computations, are of particular significance.

\begin{figure}[htbp]
    \centering
    \includegraphics[width=.8\linewidth]{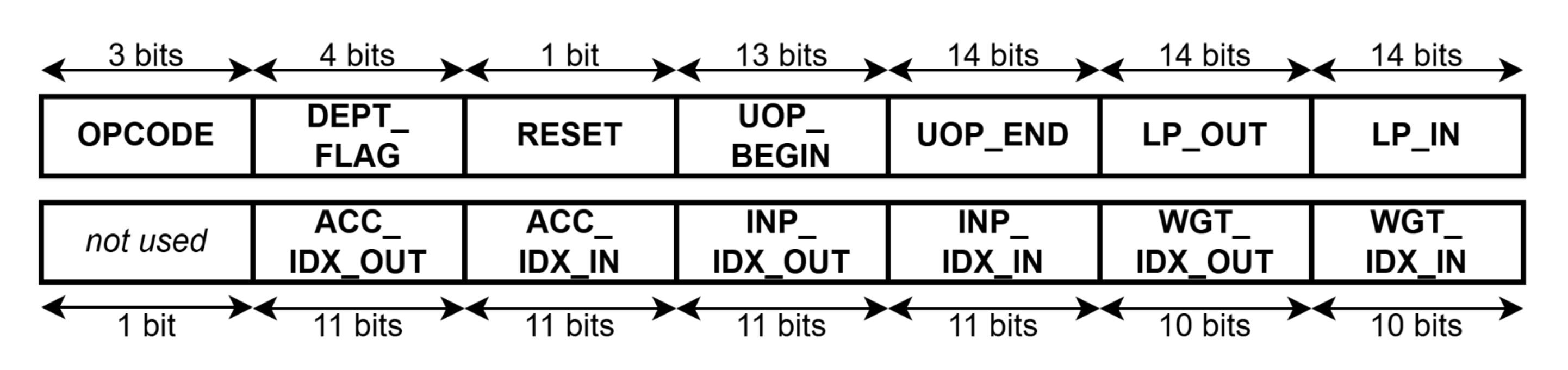}
    \caption{128-bit GeMM instruction}
    \label{fig:insn_gemm}
\end{figure}

\begin{figure}[htbp]
\centering
\includegraphics[width=0.4\linewidth]{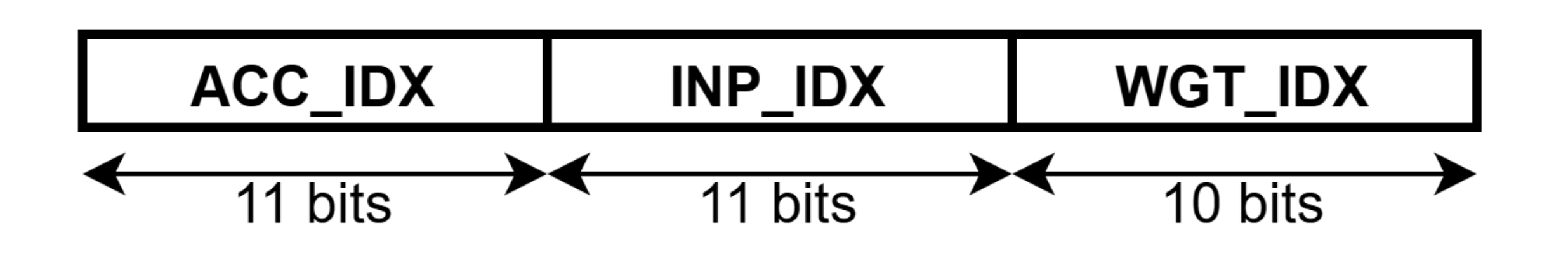}
\caption{32-bit UOP}
\label{fig:uop}
\end{figure}

\textbf{Focus on GeMM instructions:}
Figure~\ref{fig:insn_gemm} shows a GeMM instruction used by the TensorGemm module.
A GeMM (as well as an ALU) relies on a set of UOPs (micro-operations). 
These UOPs provide the initial SRAM logical addresses for ACC, INP, and WGT used by the GeMM instruction, as shown in Figure~\ref{fig:uop}.

The GeMM instruction, whose pseudo-code is given in Algorithm~\ref{listing:gemm_pseudocode}, consists of three nested loops:
(i) The outer loop (Line 1), defined by $\mathit{LP}_O$, denoted LP\_OUT in Figure~\ref{fig:insn_gemm};
(ii) The inner loop (Line 2), defined by $\mathit{LP}_I$ (LP\_IN);
(iii) The UOP loop (Line 3), defined by the number of UOPs to read, from $UOP_B$ (UOP\_BEGIN) until $\mathit{UOP}_E$ (UOP\_END);

The initial INP, WGT, and ACC addresses are provided by the UOPs (Line 4). 
These addresses are then incremented through loop iterations and instruction factors (Lines 5, 7, and 8). 
For instance, the current INP vector depends on $\mathit{INP}_O$ (INP\_IDX\_OUT) and $\mathit{INP}_I$ (INP\_IDX\_IN) as shown in Line 7.
Line 9 performs the computation $C = A \times W^T + X$. Since the TVM compiler always stores the weight matrix in its transposed form (i.e., $W = B^T$), the resulting operation is $C = A \times B + X$.

\begin{algorithm}[H]
\caption{Pseudo-code of GeMM instruction}
\label{listing:gemm_pseudocode}
\begin{algorithmic}[1]
\renewcommand{\algorithmicrequire}{\textbf{Input:}} 
\REQUIRE uop\_buf, i\_buf, w\_buf, acc\_buf 
 \FOR {$i_{out} \gets 0$ until LP$_{O}$}
   \FOR {$i_{in} \gets 0$ until LP$_{I}$}
       \FOR {$i_{uop} \gets $UOP$_{B}$ until UOP$_{E}$}
    \STATE $(i_{acc}, i_{inp}, i_{wgt}) = $ uop\_buf[$i_{uop}$]
    \STATE $x = i_{out} \times$ ACC$_O + i_{in} \times $ACC$_I + i_{acc}$
    \STATE $X = $ acc\_buf$[x]$
    \STATE $A = $ i\_buf [$i_{out} \times$ INP$_O + i_{in} \times $INP$_I + i_{inp}$]
    \STATE $W = $ w\_buf [$i_{out} \times$ WGT$_O + i_{in} \times $WGT$_I + i_{wgt}$]
    \STATE acc\_buf$[x] = A \times W^T + X$
    \ENDFOR
  \ENDFOR
 \ENDFOR 
\end{algorithmic}
\end{algorithm}

A compiler must generate adequate instructions and data memory accesses to properly configure Algorithm~\ref{listing:gemm_pseudocode} to perform the intended matrix operation. 
In the initial VTA/TVM repository, such a compiler was intertwined with the TVM compiler~\cite{tvm_github}.
Our purpose is to define an independent and stand-alone compiler. 
Our approach is detailed in the next section.

\section{Stand-alone VTA compiler}
\label{sec:compiler}
In this paper, we propose an open-source and fully-automated Python compiler pipeline from a matrix operation description down to the generation of the associated binary data and instructions, available on GitHub~\cite{onera_github}.

Our compiler adopts certain principles originating from the TVM ecosystem for VTA, namely its DRAM allocation strategy (see Section~\ref{sec:data_mapping}) and instruction sequence ordering (including reset and termination). 
Nevertheless, it was developed from scratch, guided by key objectives that significantly differentiate it from the original TVM approach.
Unlike the tightly integrated and complex transformations within TVM, our pipeline offers a more transparent path from high-level matrix operations to VTA-specific binaries.
This transparency, which is key for traceability, is achieved through a sequence of well-defined transformation steps.
Each step generates distinct intermediate outputs, making the overall transformation flow more comprehensible and verifiable.
The modular architecture also offers significant flexibility and extensibility, simplifying the exploration and evaluation of compilation choices or implementation strategies without necessitating an overhaul of the entire pipeline.
All of these contribute to meeting aeronautics certification requirements.

\subsection{Pipeline} \label{sec:automated_pipeline}
\begin{figure}[htbp]
    \centering
    \includegraphics[width=.9\linewidth]{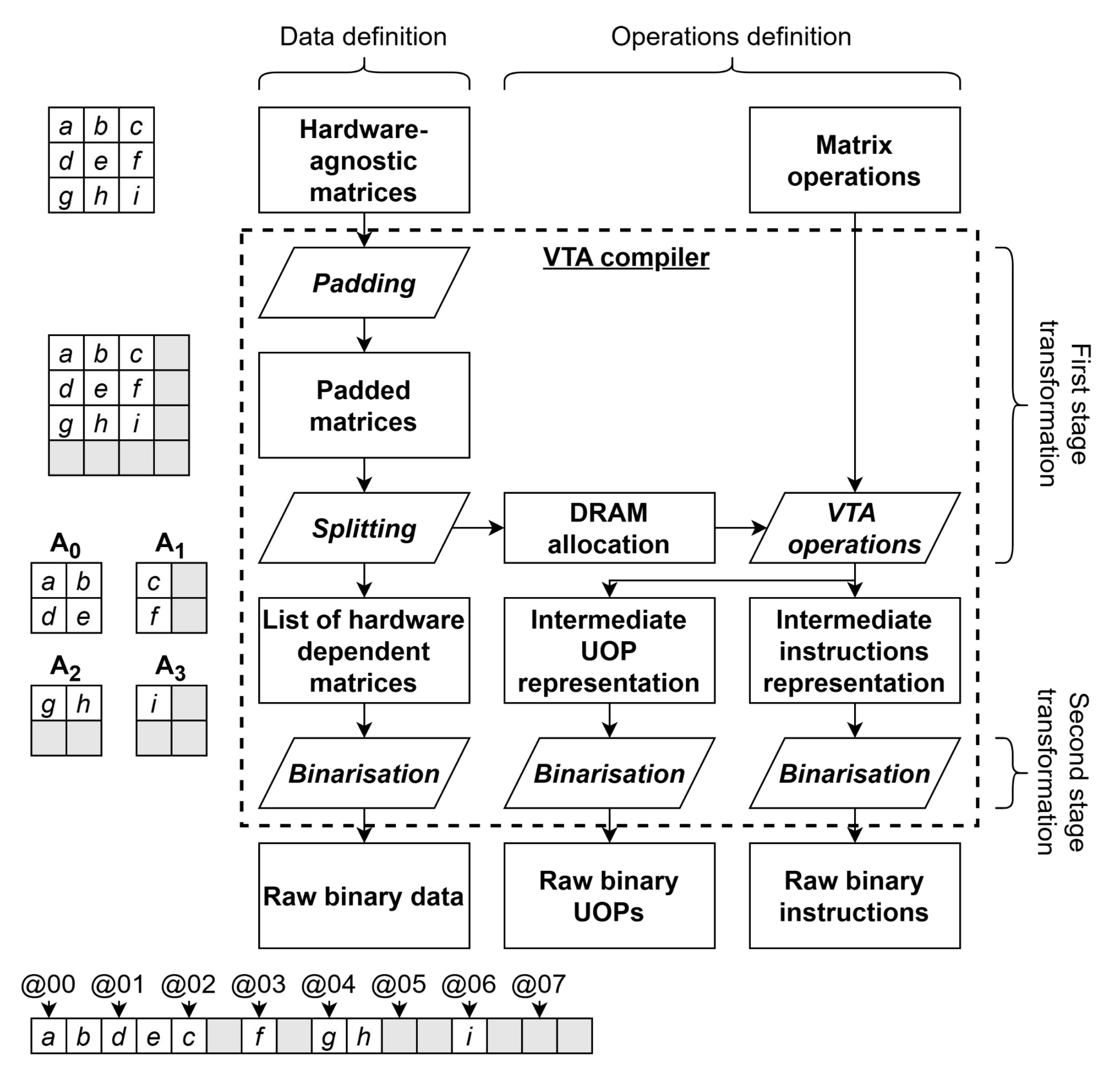}
    \caption{VTA compiler pipeline}
    \label{fig:simple_pipeline}
\end{figure}

The pipeline, shown in Figure~\ref{fig:simple_pipeline}, is divided into two parts: the data definition, which compiles the data; and the operations definition, which generates the instructions and UOPs. 
They are linked through the DRAM allocation, ensuring the correct correspondence between the data and the operations. 
Both parts are further subdivided into two transformation stages.

The VTA compiler takes as input a matrix operation and the associated matrices.
These are both hardware-independent, also referred to as hardware-agnostic.
The first stage makes the matrices and operations hardware-dependent.
Specifically for matrix multiplication, the transformation process involves converting the data into INP vectors and WGT matrices of a size compatible with the VTA (see Section~\ref{sec:VTA_architecture}), along with converting the operation into VTA instructions and UOPs.
The second stage involves encoding all preceding results into binary formats compatible with the VTA.
The compiler produces up to six binary files: \textit{input.bin} for the INP vectors, \textit{weight.bin} for the WGT matrices, \textit{expected\_out.bin} for the reference result, \textit{accumulator.bin} for the ACC vectors, \textit{instructions.bin} and \textit{uop.bin} for instructions.

\subsection{Data definition}
\label{sec:data_definition}
The data definition part is composed of three main functions: padding, splitting and binarisation.
Figure~\ref{fig:simple_pipeline} gives a simplified illustration of the data transformation process, with the \blockSize~equal to 2 (see Section~\ref{sec:VTA_architecture}).

\textbf{Padding:}
The padding function (\textit{matrix\_padding}) takes as input the initial hardware-agnostic matrix and returns a matrix, which may include additional columns and rows filled with zeros on the right and at the bottom.
The purpose is to ensure that the data conform to the VTA data structure constraints described in Section~\ref{sec:VTA_architecture}. 
Therefore, the width and height of weight matrices must be rounded up to the nearest multiple of \blockSize~(e.g., in this example, \blockSize~is 2).
However, INP and ACC are manipulated as vectors, so only the width is constrained to be the nearest higher multiple of the \blockSize.
In practice, padding INP and ACC along the height (generally) simplifies instruction generation. 
Hence, this approach is adopted throughout the subsequent paper.

In the example, let us suppose the hardware-agnostic matrix of size 3$\times$3. 
The padding function pads the matrix so that both its width and height are multiples of 2, resulting in a 4$\times$4 matrix.
The grey boxes represent the added padding.

\textbf{Splitting:}
The function \textit{matrix\_splitting} splits the padded matrix into a list of smaller matrices, called blocks, and returns the number of blocks that make up a row.
Each block is as wide as the VTA hardware \blockSize. 
All the blocks, except for the last row, must be square.
The blocks are stored in a list, ordered by row.

In the example given in Figure~\ref{fig:simple_pipeline}, the padded matrix is split into four 2$\times$2 blocks.
The resulting list of blocks is $( \block{A}{0}, \block{A}{1}, \block{A}{2}, \block{A}{3} )$.

From the VTA's perspective, WGT matrices are considered \blockSize~$\times$ \blockSize~matrices, where each matrix has a single logical address. 
Other matrices, such as INP matrices, are treated as sets of \blockSize~vectors (where each row of the matrix constitutes a vector). 
Each such vector has its own logical address. 
Consequently, an INP matrix corresponds to \blockSize~logical addresses, one for each of its \blockSize~vectors.

\textbf{Binarisation:}
Finally, the blocks are encoded into a binary file. 
The encoding follows the order in which the blocks are provided in the list, i.e., from left to right and from top to bottom.
In the case of the weight matrix, the data within the blocks are transposed (see Section~\ref{sec:VTA_instructions}); however, the order of the blocks remains unchanged.
Each initial matrix is encoded in a separate binary file, i.e., there is an INP file, a WGT file, and so forth.

In Figure~\ref{fig:simple_pipeline}, the matrix represents an INP. 
Each block is composed of \blockSize~vectors. 
Hence, each logical address refers to a row of these blocks (e.g., @00 is the first row of $\block{A}{0}$ and  @01 is the second row of $\block{A}{0}$).

\subsection{Operations definition}
\label{sec:operations_definition}
The definition of operations depends on hardware-agnostic matrix operations, the data layout specified by the data definition, and the DRAM allocation (see Section~\ref{sec:data_mapping}).
We choose to use the same DRAM allocation as the TVM compiler to serve as a reference.

The operations are generated using a template in which only the instruction fields and the UOPs are modified.
First, a pair of instructions is used to reset the VTA before starting the operations. 
This ensures that no residual data remain that could affect the execution.
Second, the data are loaded, i.e., the INP, WGT, and ACC.  
The loading capability is constrained by the size of the buffers within the SRAM; therefore, the default VTA configuration supports loading up to 2048 INP vectors, 1024 WGT matrices, and 2048 ACC vectors. 
Third, the matrix multiplication is performed using the GeMM instruction and the associated UOPs (which must be loaded).
A single GeMM instruction can perform a complete matrix multiplication provided the operand matrices fit into the SRAM.
Fourth, the element-wise operations are applied using the TensorAlu.
As with the GeMM, the corresponding UOPs are loaded, and the ALU instructions are executed.
Multiple ALU instructions can be performed sequentially.
Fifth, the resulting OUT vectors are stored back into the DRAM to be accessible by both the VTA and the CPU.
Finally, a termination sequence is executed to properly finalise the VTA execution and prepare it for the next operation.

If the data do not fit into the buffers, steps 2 to 5 must be repeated as necessary, i.e., several GeMM instructions must be executed.

\begin{figure}[htbp]
    \centering
    \includegraphics[width=.6\linewidth]{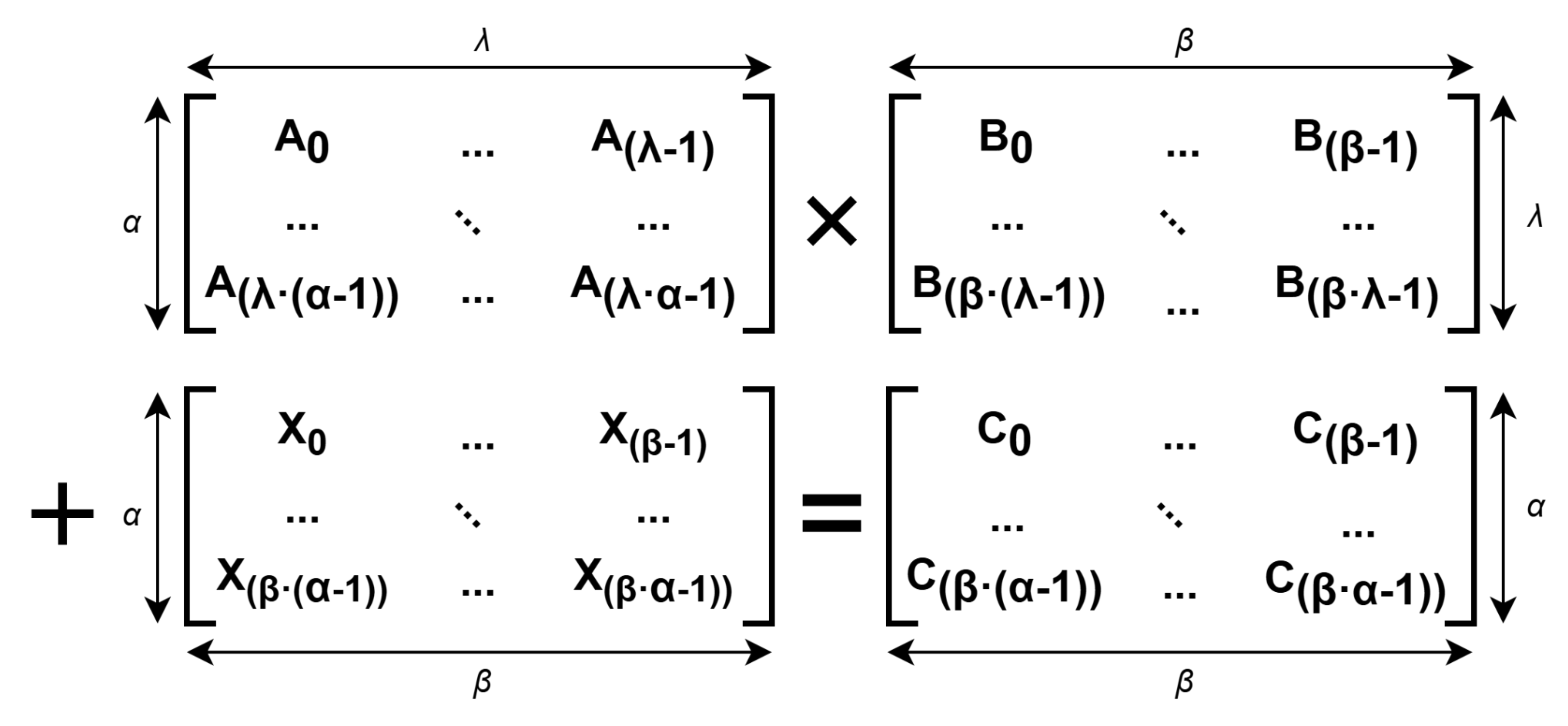}
    \caption{Block matrix multiplication}
    \label{fig:blocks_multiplication}
\end{figure}

\textbf{Focus on the GeMM instructions:}
A single GeMM instruction is capable of performing a block matrix multiplication.
Figure~\ref{fig:blocks_multiplication} illustrates such operation $C = A \times B + X$, where each block represents a \blockSize~$\times$ \blockSize~matrix. 
Definition~\ref{def:block_multiplication} provides the relationship between the blocks.

\begin{definition}[Block matrix multiplication]
    \label{def:block_multiplication}
    \itshape
    Let $\mathit{A}$ be a $\AR \times \AC$ block matrix, let $\block{A}{idx_A}$ be its blocks with $idx_A < \AC \times \AR$ the memory index.
    Let $\mathit{B}$ be a $\AC \times \BC$ block matrix, let  $\block{B}{idx_B}$ be its blocks with $idx_B < \BC \times \AC$.
    Let $\mathit{X}$, respectively $\mathit{C}$, be two $\AR \times \BC$ block matrices, let $\block{X}{idx}$, respectively $\block{C}{idx}$, be their blocks with $idx < \BC \times \AR$.
    
    $C = A \times B + X$ is equivalent to the following block matrix multiplication, $\forall i < \AR, j < \BC$:
    \[ \block{C}{i \times \BC + j} = \sum_{k=0}^{\AC-1} \left( \block{A}{i \times \AC + k} \times \block{B}{k \times \BC + j} \right) + \block{X}{i \times \BC + j} \]
\end{definition}

Note that matrix \textit{A} is the one converted into INP vectors, matrix \textit{B} into WGT matrices, and matrix \textit{X} into ACC vectors, which are loaded into the ACC buffer prior to the TensorGemm execution.
Matrix \textit{C} is the resulting matrix, corresponding to the ACC vectors after the TensorGemm execution.

\begin{figure}[htbp]
    \centering
    \includegraphics[width=.6\linewidth]{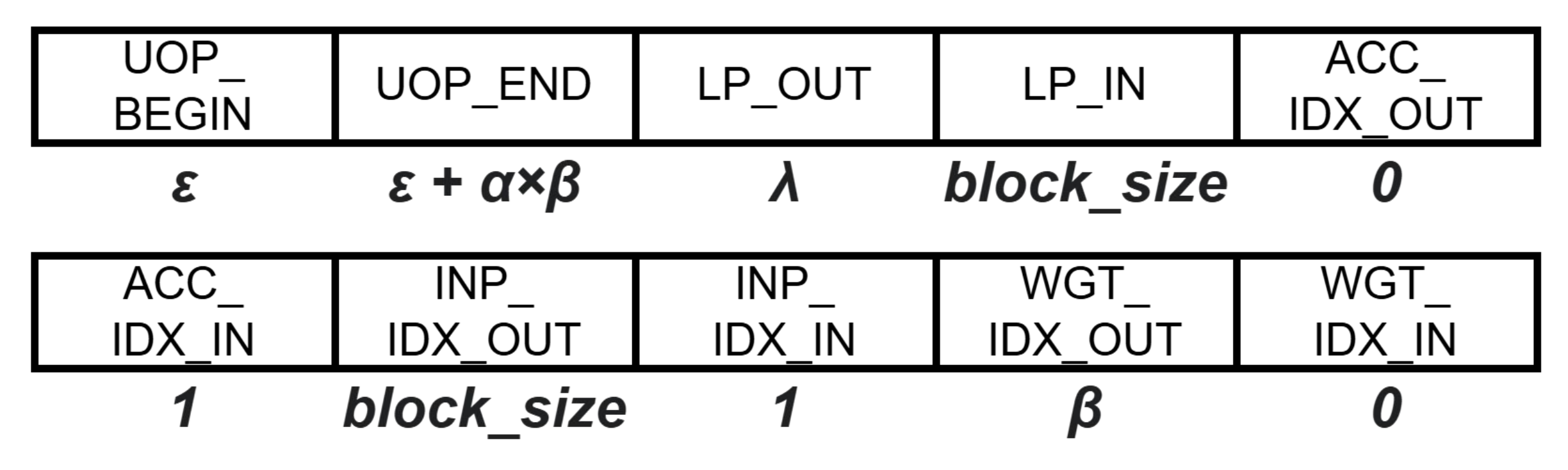}
    \caption{GeMM instruction implementation}
    \label{fig:gemm_implementation}
\end{figure}

Assuming that $\mbox{$\AR \times \AC \leq 2048$}$, $\mbox{$\AC \times \BC \leq 1024$}$ and $\mbox{$\AR \times \BC \leq 2048$}$, and that all blocks are square with a size of \textit{block\_size},
Definition~\ref{def:block_multiplication} is implemented using a single GeMM instruction shown in Figure~\ref{fig:gemm_implementation}. 
Otherwise, multiple GeMM instructions must be generated.

\begin{figure}[htbp]
    \centering
    \includegraphics[width=.6\linewidth]{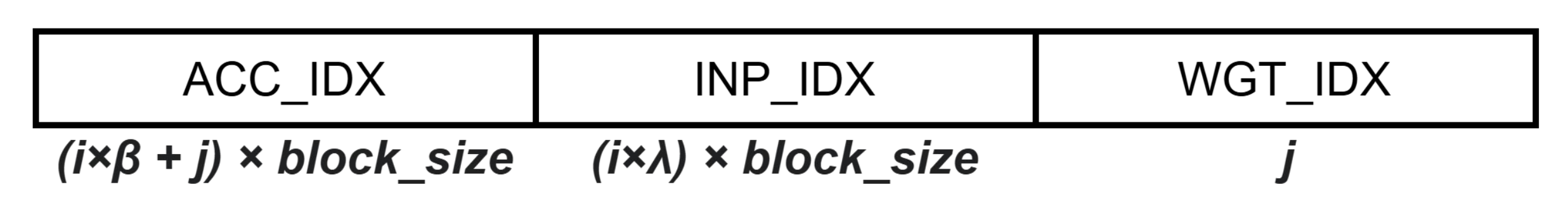}
    \caption{UOP implementation $\forall i < \AR, j < \BC$}
    \label{fig:uop_implementation}
\end{figure}

In this implementation, the inner loop (Algorithm~\ref{listing:gemm_pseudocode}, Line 2) performs the block multiplication, i.e., $\block{A}{i \times \AC + k} \times \block{B}{k \times \BC + j} + \block{X}{i \times \BC + j}$.
The outer loop (Algorithm~\ref{listing:gemm_pseudocode}, Line 1) performs the sum over the index $k$.
The UOP loop (Algorithm~\ref{listing:gemm_pseudocode}, Line 3) iterates over the $\AR \times \BC$ blocks of $C$. 
At each UOP loop iteration, a UOP is decoded (Algorithm~\ref{listing:gemm_pseudocode}, Line 4).
The first UOP decoded is the one at the address UOP\_BEGIN (here an arbitrary value $\epsilon$) in the UOP buffer, after which the address is incremented by one.
Each UOP points to the first ACC vector of a block. 
Figure~\ref{fig:uop_implementation} shows the implementation of the UOPs.

\subsection{Simple execution example}
In this example, we want to multiply the 16$\times$16 matrix \textit{A} with the 16$\times$16 matrix \textit{B} before performing a Rectified Linear Unit (ReLU) on the product.

\textbf{Data definition:}
First, \textit{A} and \textit{B} are padded if necessary. 
Here, \textit{A} and \textit{B} have a width and height equal to \blockSize, thus no padding is needed.
The matrices are then split. 
Both matrices consist of one block, respectively $\block{A}{0}$ and $\block{B}{0}$.
As explained in Section~\ref{sec:data_definition}, $\block{B}{0}$ is transposed.
Finally, the binary data are allocated as described in Section~\ref{sec:data_mapping}. 
The allocation order is as follows: the INP ($( \block{A}{0} )$), the WGT ($( \block{B^T}{0} )$), the OUT, and the UOPs. 
Their respective logical addresses in the DRAM are: @0100, @0020, @0300, and @1000.

\textbf{Operation definitions:}
First, the VTA is reset; this requires a UOP located at address @0 in the UOP buffer.

Second, the INP and WGT are loaded into their respective buffers.
$\block{A}{0}$ consists of 16 INP vectors located from @0 to @F in the INP buffer.
$\block{B^T}{0}$ consists of one WGT matrix located at @0 in the WGT buffer.

Third, the matrix multiplication implementation follows that of Figure~\ref{fig:gemm_implementation} and Figure~\ref{fig:uop_implementation}, 
with $\AR = \AC = \BC = 1$ and $\mathit{block\_size} = 16$.
Therefore, there is a single UOP associated with the operation with all its fields set to 0, which is loaded at address @1 in the UOP buffer (i.e., $\epsilon = 1$).
The parameters used by Algorithm~\ref{listing:gemm_pseudocode} to perform $A \times B$ are $\mathit{LP\_OUT} = 1$,  $\mathit{LP\_IN} = 16$,  $\mathit{UOP\_BEGIN} = 1$ and $\mathit{UOP\_END} = 2$.

Fourth, the TensorAlu submodule performs a ReLU, which is an element-wise maximum operation between each ACC vector and 0.
The ALU implementation for a ReLU is similar to the GeMM implementation.
Similarly to GeMM, this requires a UOP with all fields set to 0.

Fifth, the OUT vectors produced by the VTA, and located from address @0 to @F in the OUT buffer, are written to the DRAM at address @0300.
Sixth, the VTA execution is terminated.

\begin{figure}[htbp]
    \centering
    \includegraphics[width=.7\linewidth]{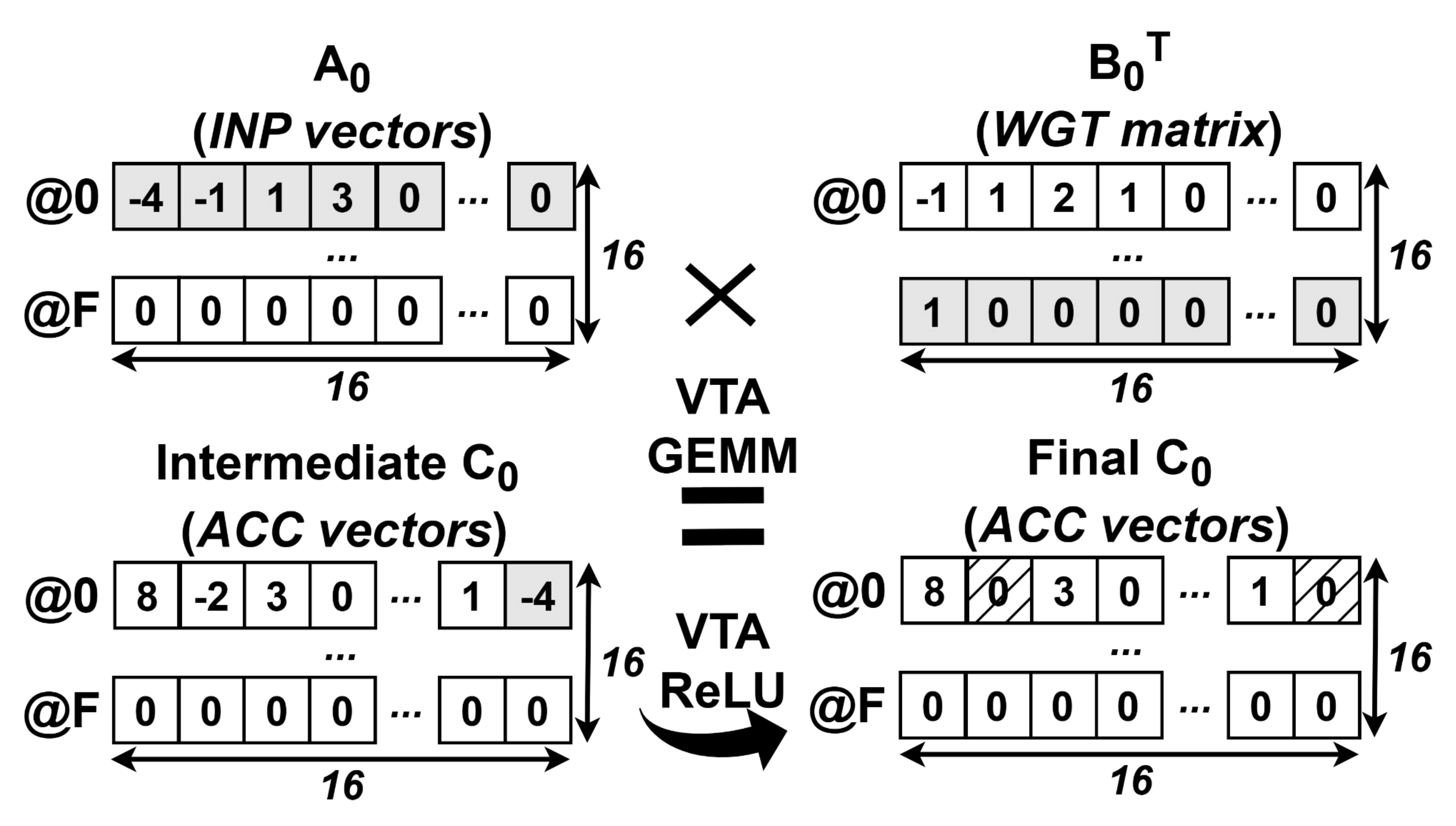}
    \caption{Matrix multiplication within the VTA SRAM}
    \label{fig:simple_16x16}
\end{figure}

\textbf{Execution:}
Figure~\ref{fig:simple_16x16} illustrates the execution of the GeMM operation. 
At each inner loop iteration (Algorithm~\ref{listing:gemm_pseudocode}, Line 2), the TensorGemm multiplies an INP vector by the WGT matrix, resulting in an ACC vector. 

The product results in 16 ACC vectors, which are then processed by the TensorAlu.
The TensorAlu applies the ReLU operation on each vector and overwrites the results in place. 
For instance, after the ReLU operation, the second and last values of ACC @0 become 0 while the others remain unchanged.

Once both the GeMM and ALU operations are completed, the OUT vectors are generated by converting the ACC vectors into signed 8-bit integers. 
The Compute module then releases the OUT buffer, allowing the Store module to write the result to the DRAM. 
Finally, the termination sequence is executed prior to the next execution.

The preceding example has demonstrated how the VTA executes a matrix multiplication (GeMM) followed by an element-wise activation function (ReLU). 
This specific sequence represents the basic building block of most layers found in CNNs.
Building upon this, Section~\ref{sec:CNN} will now explore how to effectively execute CNNs on the VTA.

\section{Execution of Convolutional Neural Networks}
\label{sec:CNN}
There exists a significant gap between CNNs and traditional matrix operations. 
CNNs are typically described in terms of tensors~\cite{tensors}, which are multidimensional data structures. 
Moreover, CNNs are composed of a sequence of operations applied to these tensors.
In this section, we outline key high-level principles of CNNs and discuss the challenges inherent in their implementation, which we will subsequently illustrate using LeNet-5~\cite{lenet5} on the VTA.
It should be noted that this paper does not explore the theoretical foundations of neural networks. 
For further details, the book~\cite{deep-learning-book} and the course~\cite{nn_math} provide comprehensive insights.

\subsection{Principles / Reminder}
In this paper, we consider two main categories of tensor operations~\cite{VTA_paper}: dense linear operations (e.g., convolution or fully connected) and non-linear operations, which include element-wise transformations (e.g., activation functions) and region-based operations (e.g., pooling). 
We define a \emph{layer} as comprising a single dense linear operation, followed by subsequent non-linear operations. 
For example, LeNet-5 is composed of five such layers. 
Its first layer consists of a convolution, followed by an activation function (e.g., sigmoid or ReLU), and then a pooling operation (e.g., average or max pooling).

\begin{figure}[htbp]
    \centering
    \includegraphics[width=.7\linewidth]{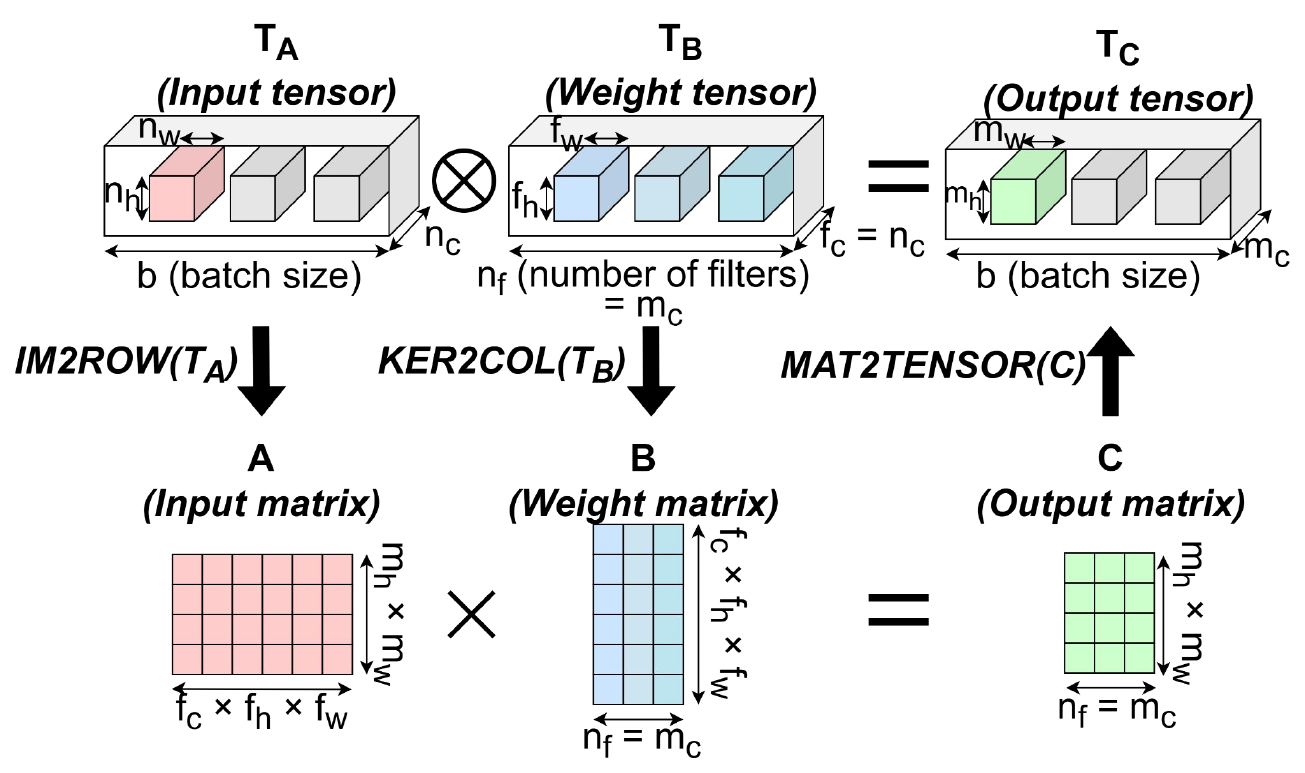}
    \caption{im2row/ker2col transformation principle}
    \label{fig:im2row_principle}
\end{figure}

\textbf{Relationship between tensor and matrix domain:} 
The operations in CNNs are typically described using 4D tensors, as shown at the top of Figure~\ref{fig:im2row_principle}, which represent collections of 3D tensors.
These tensors are commonly represented in the \emph{NCHW} format~\cite{pytorch}, where the dimensions are given in the following order:
first, the batch size ($b$ or $n_f$); 
second, the number of channels ($n_c$, $f_c$, or $m_c$); 
third, the height of the tensor ($n_h$, $f_h$, or $m_h$); 
and fourth, the width of the tensor ($n_w$, $f_w$, or $m_w$).

Input and output tensors consist of 3D tensors, processed independently. 
A batch size of 1 ($b = 1$) is used for our experiments.
Weight tensors are composed of 3D filters, each filter corresponding to one output channel.

Dense linear operations (e.g., convolution or fully connected) involve multiplying tensors together.
These operations can be transformed into matrix multiplications~\cite{fc_conv} executable by the VTA (see Section~\ref{sec:compiler}).
One such transformation method is $\mathit{im2row}$ transformation~\cite{im2row}, as illustrated in Figure~\ref{fig:im2row_principle}. 
Definition~\ref{def:im2row_principle} establishes the principles of the transformation functions. 
Once the transformations are done, the input and weight matrices are multiplied using the TensorGemm module. 
Subsequently, non-linear operations can be applied to the resulting output matrices using the TensorAlu module.

\begin{definition}[im2row, ker2col and mat2tensor principles]
    \label{def:im2row_principle}
    \itshape
    Let $T_A$ be an input tensor.
    Let $T_B$ be a weight tensor with dimensions $(n_f, f_c, f_h, f_w)$.
    Let $T_C$ be an output tensor such that $T_C = T_A \otimes T_B$, with dimensions $(1, m_c, m_h, m_w)$.
    Let $A$ be a $\AR \times \AC$ input matrix.
    Let $B$ be a $\AC \times \BC$ weight matrix.
    Let $C$ be a $\AR \times \BC$ matrix such that $C = A \times B$.
    Let $\mathit{im2row}$ be the function that converts an input tensor into an input matrix,
    $\mathit{ker2col}$ be the function that converts a weight tensor into a weight matrix,
    $\mathit{mat2tensor}$ be the transformation function that converts an output matrix into an output tensor.
    Then, if $A = \mathit{im2row}(T_A)$ and $B = \mathit{ker2col}(T_B)$,
    we have: $T_C = \mathit{mat2tensor}(C)$.
\end{definition}

\subsection{Implementation of CNNs using our VTA compiler}
\begin{figure}[htbp]
    \centering
    \includegraphics[width=0.6\linewidth]{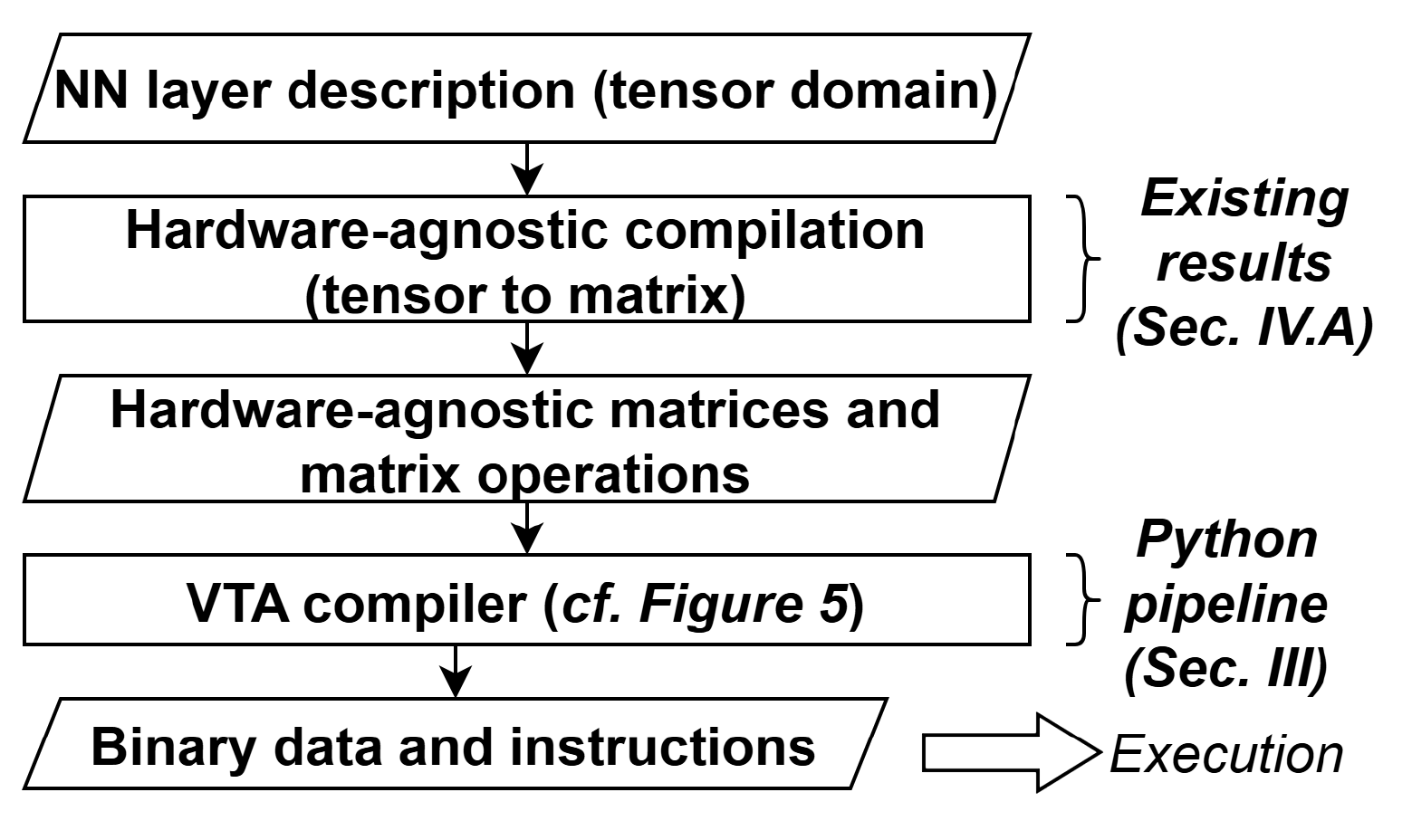}
    \caption{Compilation of one layer for the VTA}
    \label{fig:extended_pipeline}
\end{figure}

Our VTA compiler pipeline implements a single layer that is first converted into the matrix domain.
Figure~\ref{fig:extended_pipeline} shows the extended pipeline, which includes the first compilation stage responsible for the hardware-agnostic conversion of tensors into matrices. 

\textbf{Layer chaining:}
CNNs are composed of multiple layers. 
In the tensor domain, these layers are executed sequentially without specific post-processing between them. 
However, once converted into matrices through the $\mathit{im2row}$/$\mathit{ker2col}$ methods, post-processing may be required between each layer.
This post-processing typically includes reshaping the matrix to match the input format expected by the next layer.

Layer chaining is not yet automated, mainly due to the matrix reshaping and the global DRAM allocation strategy. 
The current approach to chain layers is to compile all the layers independently to generate data and instructions.
Then, the data are allocated in the DRAM and the instructions are adapted to match this allocation strategy.
Finally, the matrix reshaping is configured to get the output of a layer from the DRAM and provide the input for the next layer within the DRAM.
The matrix reshaping is performed between two VTA executions.

We chose to implement a naive solution to perform this reshaping.
It consists of the two main following stages:
(i) Convert the output matrix back into a tensor using $\mathit{mat2tensor}$;
(ii) Convert the tensor into the next input matrix using $\mathit{im2row}$.
These steps must be extended to fit the VTA implementation. 
Therefore, the first stage involves the binary decoding of the output in order to reconstruct the resulting blocks from the block matrix multiplication (see Definition~\ref{def:block_multiplication}). 
These blocks are then merged (or unsplit) to form a complete matrix, and the padding is removed. 
The resulting matrix is then converted using $\mathit{mat2tensor}$. 
The second stage consists of reapplying all the transformation steps of the extended pipeline: $\mathit{im2row}$, padding, splitting, and binarisation.

\begin{figure}[htbp]
    \centering
    \includegraphics[width=\linewidth]{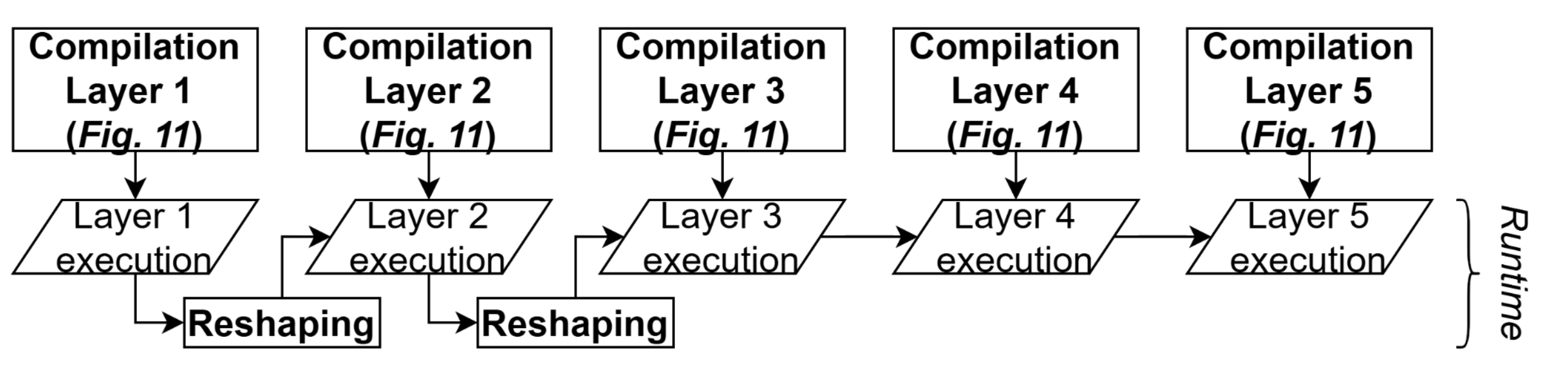}
    \caption{Layer chaining on LeNet-5}
    \label{fig:lenet5_application}
\end{figure}

Figure~\ref{fig:lenet5_application} illustrates the layer chaining on LeNet-5.

\subsection{Application to LeNet-5}
LeNet-5~\cite{lenet5} is a well-known neural network designed for handwritten digit recognition, specifically for the MNIST~\cite{mnist} dataset. 
Its architecture consists of five sequential layers: 
the first two layers combine convolution, ReLU activation, and average pooling; 
the third layer comprises convolution and ReLU; 
the fourth layer is fully connected with ReLU; 
and the final layer is a fully connected output stage producing ten classes.
LeNet-5 is applied to the VTA using the extended pipeline (depicted in Figure~\ref{fig:extended_pipeline}).
Each layer is compiled and tested independently. 
Then, the instructions and reshaping process are manually configured to combine the five layers. 
The reshaping process is performed twice: once between the first and second layers, and again between the second and third layers. 
The other layers result in a format compatible with the next one, thanks to the fully-connected layers.

\textbf{First layer:}
The input of LeNet-5 is an image in black-and-white with a resolution of 32$\times$32 pixels.
It results in an input tensor $(1, 1, 32, 32)$.
The $\mathit{im2row}$ transformation results in a 784$\times$25 input matrix.
This matrix is padded and split to result in a block matrix with $\AR = 49$ and $\AC = 2$, where each block is 16$\times$16.
Then, it is encoded into binaries and executed on the VTA.  
The output is subsequently decoded into a hardware-agnostic matrix of size 196$\times$6, and then converted into an output tensor of shape $(1, 6, 14, 14)$.

\textbf{Other layers:}
All the layers are compiled prior to executing LeNet-5 in the same way as layer 1.
All of these binaries, along with those related to the subsequent layers, are allocated in DRAM and must be taken into account when defining the logical addresses within the instructions. 
Consequently, only the conversion of the previous output tensor into binaries (i.e., the second stage of reshaping) is performed between two VTA executions, directly on the host machine.

\section{Simulation and results}
\label{sec:simulation}
The VTA/TVM project from~\cite{tvm_github} provides two simulators: a functional simulator and a cycle-accurate simulator. 
However, these simulators were not easily executable, as the functional simulator was embedded within the compilation pipeline, while the cycle-accurate simulator did not accept the binary files resulting from the compilation. 
To address this, we extracted and modified both simulators to make them more accessible, ensuring that they both accept raw binaries as input. 

We have compiled and simulated the LeNet-5 model on both VTA simulators.
The PyTorch model~\cite{pytorch} served as the reference for the simulation.

\subsection{Functional simulator}
The functional simulator, written in C++, provides fast results strictly equivalent to those produced by the VTA hardware implementation.
Only limited execution details can be observed with this simulator:
the total size of data exchanged with DRAM, the total number of loops, and the execution output.

\textbf{LeNet-5 simulation:}
This simulator returns the bit-accurate result almost instantaneously, thereby proving the correctness of the compilation.
It also reports that the execution requires 2942 GeMM loops to perform all the matrix multiplications required for LeNet-5.

\subsection{Cycle-accurate simulator}
The cycle-accurate simulator adheres to the hardware constraints as it is based on CHISEL (Constructing Hardware In a Scala Embedded Language)~\cite{chisel}, a Hardware Description Language (HDL). 
Furthermore, the CHISEL language makes the simulator easy to modify, allowing for the customisation of the execution report to highlight or adapt specific details.

We limited the cycle-accurate simulation to the Compute module, as it constitutes the core of the execution and was the primary focus of our analysis.

\textbf{LeNet-5 simulation:}
This simulation takes significantly longer, approximately 4 hours.
Like the functional simulator, it produces the bit-accurate result.
Additionally, it provides cycle-level details, including the current INP vector and WGT matrix being used, as well as the ACC vector produced. 
The simulation indicates that the VTA takes 2972 cycles to execute these TensorGemm operations, including decoding the GeMM instruction and checking buffer availability. 
Therefore, the VTA is able to almost complete an entire GeMM loop in each cycle. 
In this case, with $\blockSize = 16$, a GeMM loop consists of 256 Multiply-Accumulate (MAC) operations. 
Considering a modern SIMD CPU capable of performing 16 MACs per cycle~\cite{cpu}, at least 47552 total cycles would be required (approximately 16 times more cycles per loop compared to the VTA). 
In total, 6358 cycles are needed to execute LeNet-5 (without considering Load and Store operations). 
Considering an FPGA with a clock speed of 650~MHz, LeNet-5 is executed in 9.8 microseconds.
The considered CPU must have a frequency of approximately 10~GHz to achieve the same execution time.
This frequency is significantly beyond the typical operating range of current commercial CPUs.

\section{Related work}
\label{sec:related_work}

\subsection{Hardware accelerators}
Hardware accelerators are essential for executing modern Machine Learning algorithms, such as CNNs. 
They aim to efficiently execute dense linear algebra operations~\cite{hw_sw_survey,cnn_hw_survey}, the most computationally critical parts of CNNs.
A wide variety of hardware accelerators exists, including Tensor Processing Units (TPUs)~\cite{tpu_matrix}, Neural Processing Units (NPUs)~\cite{npu_example}, Graphics Processing Units (GPUs)~\cite{gpu_example}, and FPGA-based accelerators~\cite{VTA_paper}.
Most feature parallel architectures, which are well-suited for executing matrix-based operations.

When selecting a hardware accelerator, two major factors must be considered: performance and flexibility.
TPUs generally offer the highest performance, whereas FPGA-based accelerators provide the greatest flexibility~\cite{hw_comparison}.
For safety-critical applications, additional considerations should be taken into account, such as the availability of detailed internal architectural information.

We selected the VTA (Versatile Tensor Accelerator)~\cite{VTA_paper} for this study. 
Its open-source description using CHISEL, an RTL-like description language, provides fine-grained hardware details that are key to hardware analysis and formal verification in safety-critical applications.
Leveraging the Scala programming environment, CHISEL facilitates the setup of test benches and hardware simulations. 
This environment also provides CHISEL with efficient simulation-based verification capabilities~\cite{chiselverify}.
Moreover, it offers formal verification features~\cite{chiseltest}, valuable for safety-critical analyses.

\subsection{Compiling tensors into matrices}
Executing tensor-based NN operations on matrix accelerators like VTA requires transforming tensors into suitable matrix representations. 
Compilers like TVM offer various techniques for this purpose~\cite{tvm}.
Two common approaches are tensorisation and explicit matrix conversion methods like $\mathit{im2col}$/$\mathit{im2row}$.

Tensorisation~\cite{tvm} maps high-level tensor operations directly onto VTA's architecture for execution. 
While potentially efficient, this often involves complex sequences of transformations within the compiler flow~\cite{tvm_docs}.

Alternatively, transformations like $\mathit{im2col}$~\cite{im2col} (image-to-column) and $\mathit{im2row}$~\cite{im2row} (image-to-row) explicitly restructure input tensors (e.g., from convolutional layers) into large matrices suitable for standard GeMM operations. 
These methods are conceptually similar, differing mainly in the output matrix layout (transpose relationship). 
A known drawback is the significant memory overhead due to data replication within the generated matrix. 
However, their transformation logic is often considered more direct compared to multi-step tensorisation pipelines. 
Research efforts focus on optimising these methods to mitigate the memory overhead~\cite{acetone_extended}.

Other techniques, such as channel merging and kernel interleaving~\cite{mbw_mac}, target specific hardware (e.g., Multi-Bit-Width units) to potentially avoid explicit reshaping between some layers, though possibly at the cost of less standard data layouts.

In our extended pipeline, we selected $\mathit{im2row}$ partly because its resulting matrix layout appears well-suited to the vector processing architecture of VTA's TensorAlu module, facilitating efficient mapping of computations.

\subsection{Certification aspects}
Applying ML accelerators like VTA in safety-critical domains, such as avionics, requires stringent certification regarding predictability, traceability, and verifiable semantic preservation.
These certification requirements are anticipated based on the EASA (European Union Aviation Safety Agency) concept paper~\cite{easa_concept_paper}, which will define the regulatory framework for ML in critical avionics.
The joint international working group, EUROCAE WG-114 and SAE G-34, gathering industry, academia, and regulatory authorities (including EASA), is developing the ED-324/ARP6983 standard.
This standard is expected to be recognised as a Means of Compliance (MOC) with the upcoming regulatory framework, meaning its application will ensure compliance.

The ED-324/ARP6983 standard promotes a design assurance process based on a W-shaped development lifecycle.
The first 'V' is dedicated to the design and validation. 
The second 'V' focuses on the implementation and verification on the target hardware.
A primary purpose of the standard is to capitalise extensively on existing and well-established avionics guidelines, such as DO-178C~\cite{DO178C} for software development and DO-254~\cite{DO254} for hardware development.
Adherence to DO-254 and DO-178C is therefore a key consideration in the development of the VTA compiler.

\section{Conclusion}
This paper addressed significant challenges hindering the adoption of the VTA in computationally intensive fields, particularly safety-critical systems such as aeronautics. 
To tackle these issues, we developed a fully automated Python compiler pipeline, transforming hardware-agnostic matrix operations into VTA-compatible binary data and instructions. 
We detailed the processes of data padding, splitting, and binarisation, along with the generation of VTA instructions, including the critical GeMM operations for matrix multiplication. 
Furthermore, we outlined the principles for executing CNNs on the VTA by converting tensor operations into matrix multiplications using methods such as $\mathit{im2row}$/$\mathit{ker2col}$, applying this methodology to the LeNet-5 network. 
We also provided a simulation framework to investigate different compilation strategies.

Future work can effectively build upon the compiler and simulation framework presented. 
First, this framework provides the tools to investigate, simulate, and compare different compilation strategies and transformation techniques for the VTA. 
For instance, it can evaluate advanced methods, such as the one proposed in~\cite{mbw_mac} for directly chaining multiple network layers, assessing their performance and correctness implications. 
Second, once compilation strategies are clearly identified and validated through this framework, the insights gained will enable further automatising the compiler pipeline for streamlined and direct implementation of NNs on the VTA. 
Finally, complementing these compilation-focused efforts, a crucial perspective involves the detailed analysis and formal verification of the VTA architecture itself. 
This could leverage the VTA's hardware description (e.g., in CHISEL) to provide stronger guarantees, paramount for safety-critical applications.

\section*{Acknowledgment}
This work has benefited from the AI cluster ANITI2 and from the PHYLOG 2 pro-
ject funded by the French government through the France Relance program, based on the funding from
and by the European Union through the NextGenerationEU program.

\bibliographystyle{IEEEtran}
\bibliography{IEEEabrv,reference}

\end{document}